\documentclass[sigplan,screen]{acmart}
\setcopyright{none}
\renewcommand\footnotetextcopyrightpermission[1]{}
\usepackage{enumerate}
\usepackage{color}
\usepackage{neuralnetwork}
\usepackage{tikz}
\usepackage{diagbox}
\usepackage{amsmath}

\newcommand{\summation}[2]{\underset{#1}{\overset{#2}{\sum}}}
\newcommand{\pai}[2]{\underset{#1}{\overset{#2}{\Pi}}}
\usetikzlibrary{er,positioning}
\AtBeginDocument{%
  \providecommand\BibTeX{{%
    \normalfont B\kern-0.5em{\scshape i\kern-0.25em b}\kern-0.8em\TeX}}}

\begin{document}
\title{Survey of Query-Based Text Summarization}

\author{Hang Yu}
\affiliation{%
  \institution{CS @ UIUC}
  \city{Urbana}
  \state{IL}
  \country{USA}
  \postcode{61801}
}
\email{hangy6@illinois.edu}

\author{Jiawei Han}
\affiliation{%
  \institution{CS @ UIUC}
  \city{Urbana}
  \state{IL}
  \country{USA}
  \postcode{61801}
}
\email{hanj@illinois.edu}

\renewcommand{\shortauthors}{Trovato and Tobin, et al.}
\newcommand{\tbd}[0]{\color{red}To be finished!\color{black}}
\newcommand{\tbc}[0]{\color{red}cite needed!\color{black}}
\newcommand{\m}[0]{\color{red}MARK\color{black}}

\begin{abstract}
   Query-based text summarization is an important real world problem that requires to condense the prolix text data into a summary under the guidance of the query information provided by users. The topic has been studied for a long time and there are many existing interesting research related to query-based text summarization. Yet much of the work is not systematically surveyed. This survey aims at summarizing some interesting work in query-based text summarization methods as well as related generic text summarization methods. Not all taxonomies in this paper exist the related work to the best of our knowledge and some analysis will be presented.
\end{abstract}


\maketitle
\pagestyle{empty}

\section{INTRODUCTION}
Text summarization problems are palpable in various real-world applications. With the development of the internet, information nowadays spreads at a speed higher than ever. Amid information spread on the internet, a very large portion of the information is shared in text. It can be an arduous work for users to summarize and identify useful information from this huge amount of text data. Thus automatic text summariztion will efficiently summarize the text. Much effort has been made to summarize the document in general, however, in real world problems, users may only be interested in a certain aspect of the information in the document which is not the major part of the document. Thus insisting on general text summarization may result in the summary that does not contain information that the users are interested in. Therefore, query-based text summarization which uses user defined query information as guidance to interfere with the automatic text summarization process.

Query-based, also named \textbf{topic-based}, \textbf{user-focused} or \textbf{query-focused}, text summarization can be regarded as a text summarization task that leverages the query information provided by the users.  For generic text summarization tasks, the aim is to obtain a generic summary or abstract of given documents. The response to the query extracts or generates text that both covers query-related key points in the documents and compresses the documents. In this sense, text summarization can be regarded as a higher hierarchy with three subclasses: generic text summarization, factoid text summarization (aka question-answering) and query-based text summarization. But sometimes, the question answering is also regarded as a type of query-based summarization since there are many studies that do not force the model to reply to the question with a simple factoid text unit. In some cases, the reply to the questions is a summary of the question related content. And we will also introduce such kind of question-answering papers in the later sections.

From the perspective of text summarization, query-based text summarization can be divided into query-based extractive and abstractive text summarization. Considering the user-provided query information, query-based text summarization can be divided into word query and sentence query. Also, some studies on this topic focus on single document summarization (SDS) and some studies focus on multi-document summarization (MDS)\cite{rahman2015survey}. Based on these dichotomies, many query-based text summarization surveys have been conducted. In \cite{nenkova2012survey}, generic text summarization techniques are surveyed and in \cite{gupta2010survey, moratanch2017survey}, extractive text summarization techniques are surveyed. Although both survey papers mentioned the query-based text summarization, they do not expand enough information on this specific topic. Afantenos et al.\cite{afantenos2005summarization} surveyed on text summarization about medical documents. Although this is a specific aspect of text summarization, the query information is confined to medical related only. Damova et al.\cite{damova2010query} surveyed query-based text summarization in 2010, but it turns out that during the following decade, with the emerging of Word2Vec\cite{mikolov2013distributed}, end-to-end sequence generation models\cite{cho2014learning, sutskever2014sequence} and various attention mechanism\cite{bahdanau2014neural, luong2015effective, yang2016hierarchical, vaswani2017attention}, query-based text summarization techniques have gained many advances along with text summarization. Thus this survey will focus on synthesizing specific techniques in query-based text summarization and make up for the gap that the most recently proceeding in this topic are not comprehensively surveyed.

\section{OVERVIEW}
This part is an overview of the structure of the survey. The first dichotomy we will dive into is the extractive techniques and abstractive techniques. These two types of text summarization are different in nature since extractive techniques focus on salient parts in the document and directly extract the salient discourse from the document. Hence query-based extractive summarization is most likely to score on the salience of the discourse with respect to the query dimension. In contrast, abstractive text summarization focuses on identifying the salience of the information in the text and generating new text that concludes the information identified, and the new text is not seen in the document.
Although most studies in text summarization are based on extractive or abstractive text summarization, there are some studies attempting to fuse these two kinds of techniques and use hybrid methods. We will also discuss some major findings in these works after the extractive and abstractive part.

In either query-based extractive or abstractive text summarization section, we will majorly introduce the related work according to 2 basic compartments of machine learning techniques: \textbf{unsupervised learning, supervised learning}. By synthesizing different papers in each compartment, common features will be extracted to both conclude the core idea of this compartment and provide insight for further research.

\section{Text Summarization Evaluation}
One of the most important portions in developing an effective machine learning algorithm is to have validated evaluation that ensures the generalization of the developed algorithm or method. Specifically, evaluation is composed of two parts, evaluation datasets and evaluation metrics. For different machine learning tasks, different evaluation datasets and metrics should be chosen for the correct generalization purpose. So, before the summarization techniques are introduced, one important question to be answered for query-based text summarization is how should the summaries predicted by the model be evaluated?

\subsection{Evaluation Datasets}
In some early studies, human experts evaluations were used to measure the effectiveness of the model or the method. However, much more complicated methods are proposed with exponential growth and they need much larger dataset to both train and test their effectiveness. Hence, human experts evaluations can be much less than the requirement. Thus, many datasets targeting to solve this problem have been developed.

The most obvious sources that might be served as text summarization datasets are news corpora because they are sorted and archived systematically and are very large-scale datasets, and, most importantly, they typically have headlines or abstracts written by experts that summarizes the news which is a good reference source. For example, NYT corpus\cite{sandhaus2008new} collects news and articles from the New York Times magazine with experts summarized abstracts and CNN/DailyMail dataset\cite{hermann2015teaching} is composed of the articles that have highlights made by the authors. These two datasets are the most prevalent datasets in generic text summarization tasks, however, both the abstracts and highlights in the news and articles of these two datasets are summaries of the generic gist rather than specific facets. Thus, directly applying these two well-developed datasets to query-based text summarization tasks is not feasible.

For most studies about query-based text summarization, DUC (Document Understanding Conferences), conducted annually by NIST (National Institute of Standards and Technology), datasets are typically selected as evaluation datasets. DUC 2005 and DUC 2006 datasets are regarded as standard datasets used for query-based summarization methods evaluation\cite{litvak-vanetik-2017-query}. The reason is that the DUC 2005 dataset has 50 document sets and for every document set a short query which is typically 1-3 sentences is applied. Also, four to nine gold references are supplied for each document set. Correspondingly, DUC 2006 also has 50 document sets and each document set has short query and four gold references. More importantly, each document set is from a different topic with multiple topic related documents inside.

\subsection{Evaluation Metrics}
The answer to what is a good metric for text summarization tasks is not trivial because the actual quality of the text summaries is much more subjective than label matching in discriminative models such as text classification models, i.e. good performance on quantitative analysis is not necessarily equivalent to good performance on qualitative analysis. In general, many scholars studying text summarization are prone to regard summaries from human experts as golden references. However, it turns out that manually checking summaries generated by the models can be both expensive and inefficient, hence the goal of automatic text summarization evaluation task is to find rules or metrics to be applied to measure the quality of summaries made by the text summarization models as much like human evaluation as possible.

Some studies attempted to convert the evaluation of summaries to a reference matching problem that can be measured by precision, recall and F1-score, for instance in extractive summarization evaluation, some studies tried to compare a set of extracted sentences to a set of "correctly" extracted sentences. On top of that, Saggion et al.\cite{saggion2002meta} further proposed three content-based methods that measure similarity between the references and summaries. However, as Lin\cite{lin2004rouge} pointed, those content-based methods did not show the correlation between the automatic evaluation methods and human judgements. Then Lin further proposed an automatic summarization evaluation metric, ROUGE. Among the proposed methods, ROUGE-2, ROUGE-L, ROUGE-W and ROUGE-S worked well in single document text summarization task; ROUGE-1, ROUGE-L, ROUGE-W, ROUGE-SU4, ROUGE-SU9 had good performance in evaluating short text summaries. Justifications were presented in this paper about that high correlation with human judgements was achieved in single document summarization evaluation tasks but in multi-document summarization tasks, this topic is yet to be explored.

\section{EXTRACTIVE TECHNIQUES}
\subsection{Unsupervised Query-Based Summarization}
Conroy et al.\cite{conroy2006topic} proposed an query-based extractive text summarization that used unsupervised learning. One common practice in extractive text summarization is to score each text unit, in these paper sentences, with an oracle score and then select the text units with high scores as part of the summary. In this paper, however, the oracle score is intended for the appropriateness of a sentence to be included in the summary for topic $\tau$. The true objective oracle score is defined in the paper as:
\begin{align*}
    w(x) = \frac{1}{|x|}\summation{t\in T}{}x(t)P(t|\tau)
\end{align*}
where |x| is the number of distinct terms sentence x contains, T is the set of all terms in topic $\tau$, x(t) is the indicator function of whether term t is in sentence x or not, and $P(t|\tau)$ is the probability that a human will choose term t in a summary given a topic $\tau$. The maximum likelihood approximation is used to approximate $P(t|\tau)$ as:
\begin{align*}
    \hat{P(t|\tau)} = \frac{1}{h}\summation{i=1}{h}c_{it}(\tau)
\end{align*}
where $c_{it}(\tau)$ is the indicator function of whether the i-th summary contains the term t. Further, to work out the maximum likelihood estimation in an analogous way, a set of terms is isolated where \textit{query terms} are from gleaned from topic description and \textit{signature terms} are derive from the collection of documents related to the topic. In this way, the analogous fashion of the approximation is written as:
\begin{align*}
    \hat{P(t|\tau)} = \frac{1}{2}q_t(\tau) + \frac{1}{2}s_t(\tau)
\end{align*}
where $q_t(\tau) = 1$ if t is a query term
for topic $\tau$ and 0 otherwise and $s_t(\tau) = 1$ if t is a signature term for topic $\tau$
and 0 otherwise.

For both the oracle score and the approximation, the summary consists of sentences with at least 8 distinct terms and with top scores. The hard coded threshold 8 is selected according to their previous analysis of the sentence splitter. DUC05 dataset where there are summaries that are both marked general or specific and the ROUGE-2 score is used as the metric for this study.

The work in Controy et al. was focusing on the single document query-based text summarization. However, many applications also require multi-document query-based summarization. Jade et al.\cite{goldstein2000multi} proposed four major differences between the single-document and multi-document text summarization:
\begin{enumerate}[1.]
    \item \textbf{Information redundency.} In multi-document query-based text summarization, a group of documents from the same topic can produce a much higher level of information redundancy than a single document in that certain topic.
    \item \textbf{Temporal dimension.} Some types of text data such as news reports will have the issue to be overridden by the later update and sometimes they might be packed together in the dataset.
    \item \textbf{Compression ratio.} For a specific topic, the information in the summary is limited, hence typically the more documents in a topic to be summarized for a multi-document summarization, the more compression will be required.
    \item \textbf{Correference problem.} Baldwin et al.\cite{baldwin1998dynamic} pointed out that the correference problem in the multi-document summarization is more challenging than that in the single-document summarzation.
\end{enumerate}

Based on the four significant differences, they enhanced their previous work on "Maximal Marginal Relevance" (MMR) by proposing a new metric that is effective to reduce redundancy and they call the metric as Maximal Marginal Relevance Multi-Document metric (MMR-MD). The formulus of the metric is defined as follow:
\begin{align*}
    MMR-MD = &\underset{P_{ij}\in S}{argmax}[\lambda Sim_1(P_{ij}, Q, C_{ij}, D_i, D)\\
    &- (1 - \lambda)Sim_2(P_{ij}, P_{nm}, C, S, D_i)]\\
    Sim_1(P_{ij}, Q, C_{ij}, D_i, D) = &w_1*(P_{ij}\cdot Q) + w_2*coverage(P_{ij}, C_{ij})\\
    &+ w_3*content(P_{ij})\\
    &+ w_4*time_sequence(D_i, D)\\
    Sim_2(P_{ij}, P_{nm}, C, S, D_i) = &w_a*(P_{ij}\cdot P_{nm})\\
    &+ w_b*clusters_selected(C_{ij}, S)\\
    &+ w_c*documents_selected(D_i, S)\\
    coverage(P_{ij}, C) = &\summation{k\in C_{ij}}{}w_k*|k|\\
    content(P_{ij}) = &\summation{W\in P_{ij}}{}w_{type}(W)\\
    time\_sequence(D_i, D) = &\frac{timestamp(D_{maxtime}) - timestamp(D_i)}{timestamp(D_{maxtime}) - timestamp(D_{mintime})}\\\\
    clusters\_selected&(C_{ij}, S) = |C_{ij} \cap \underset{v, w; P_{vw}\in S}{\cup} C_{vw}|\\
    documents\_selected&(D_i, S) = \frac{1}{|D_i|}*\summation{w}{}P_{iw}\in S
\end{align*}
The original paper has a detailed explanation for each symbol in the formulas and the rationale behind the formulas of coverage, content, time sequence, clusters selected, and documents selected can be found in section 3 of the original paper. The high level idea of their method is contained in the first three formulas here. $Sim_1$ and $Sim_2$ are computed based on the detailed indexes mentioned above and $Sim_1$ computes the similarity metric of the relevance ranking for query-based summarization purpose while $Sim_2$ computes the anti-redundancy metric that help to filter out the redundancy information. By leveraging the query-based summarization metric and the anti-redundancy metric through a tunable hyperparameter $\lambda$ and maximizing the objective, the query-based summary from multiple documents is extracted.

Although very few studies were conducted in unsupervised query-based text summarization, there are sufficient amounts of papers demonstrating unsupervised generic text summarization. Natalie et al.\cite{schluter2015unsupervised} proposed an unsupervised learning extractive text summarization method using coverage maximization with syntactic and semantic concepts. The focus of the paper is to argue that coverage maximization with bigram concepts is better than that with syntactic and semantic concepts. As demonstrated in their work, the bigram model has to make some preprocessing such as stop-word removing and word stemming whereas these preprocessing are not necessary using syntactic dependencies or semantic frame. Similar to some query-based text summarization work aforementioned, much effort of their work is put on the text analysis, and the features to be used for sentence extraction are inevitably related to some word count. Hard assignment of some thresholds, such concept count cut-off in this work and distinct terms cut-off in a selected sentence for the summary in Conroy et al.\cite{conroy2006topic}, can still be problematic in many cases.

Later improvements majorly focused on graph-based methods where sentences are converted to nodes and the edge weights are measured by sentence similarity until the large pre-trained transformers, such as BERT, are proposed by Devlin et al.\cite{devlin2018bert} in natural language processing. BERT applied an unsupervised learning method to build the pre-trained language model. The pre-trained language model can then be fine-tuned to finish many downstream tasks. This is similar to a common practice in computer vision tasks called transfer learning. Original BERT paper used word masks to mask out the word in a sentence and trained by predicting the mask. Since the unlabeled text data is typically abundant, the model can be trained to capture a large amount of information without manually setting rules for the natural language. Thus, with the information captured in the unlabeled data, the pre-trained model can be easily fine-tuned to fit for the specific downstream task and Devlin et al. has already proved that this pre-training method helps to outperform state-of-the-art in 11 natural language tasks.

Similar to the BERT, Zhang et al.\cite{zhang-etal-2019-hibert} pre-trained the model by masking out the sentences in documents instead of the words and then predicted it to train the pre-trained model. Instead of using a single stack of encoders in their proposed architecture, they experimented with a hierarchical representation with both sentence encoders and document encoders. The sentence encoders encoded the sentence word by word and then the document encoders encoded the document sentence by sentence. The pre-trained model, which is called Hierarchical Bidirectional Encoder Representations (HIBERT) in the paper, is based on unsupervised learning, however, for the specific text summarization task, Zhang et al. still used supervised learning extractive summarization techniques. However, later Xu et al.\cite{xu2020unsupervised} argued that the self-attention scores in the sentence-level transformers of the pre-trained HIBERT has already become meaningful for estimating the salience of the sentences in the documents.

For the summarization part, Xu et al. proposed two ranking criteria and leverage the two criteria through the tunable parameters to train the model. The first criteria is based on the probability of sentences in a document. They hypothesize that:
\begin{align*}
    P(\mathcal{D}) = \pai{i=1}{|\mathcal{D}|}P(S_i|S_{1:i-1}) \approx \pai{i=1}{\mathcal{D}}P(S_i|\mathcal{D}_{\lnot S_i})
\end{align*}
where $P(S_i|\mathcal{D}_{\lnot S_i})$ represents the probability that all the masks in the document are replaced by the sentence $S_i$. The approximation is applied here due to the difficulty of estimating the original probability in a bidirectional sequence model. Then the score is computed and normalized by the different length of the sentences to make their scores comparable as:
\begin{align*}
    \hat{r}_i = \frac{1}{|S_i|}\summation{j=1}{|S_i|}P(w_j^i|w^i_{0:j-1}, \mathcal{D}_{\lnot S_i})
\end{align*}
where $w_i^j$ represents the j-th word in the i-th sentence. And finally the first ranking criteria is computed by normalizing $\hat{r}_i$ across sentences in a document as:
\begin{align*}
    \Tilde{r}_i = \frac{\hat{r}_i}{\sum_{j=1}^{|D|}\hat{r}_j}
\end{align*}

The second ranking criteria is applied to model the contributions of other sentences to the current sentence. Using the self-attention matrix $\textbf{A}$, the second ranking score for sentence $S_i$ can be derived from the previously computed first ranking score as follow:
\begin{align*}
    r'_i = \summation{j=1, j\neq i}{|\mathcal{D}|}\textbf{A}_{i, j}\times \Tilde{r}_j
\end{align*}
Finally the overall ranking score for extractive summarization is defined as:
\begin{align*}
    r_i = \gamma_1\Tilde{r}_i + \gamma_2r'_i
\end{align*}

From the synthesized paper above and to the best of our knowledge, unsupervised extractive text summarization was at first more focused on rule based natural language analysis. Correspondingly, unsupervised query-based extractive text summarization typically follows the same rule and the most significant difference between query-based and generic summarization is that the score for topic relevance. A significant disadvantage is that to set the appropriate rules for the summarizers, researchers need to do some complicated natural language analysis first. This preprocessing procedure can be prohibitively expensive when applied to very large-scale corpus data. Also, there are chances that many of the text data are noisy and the rules are also affected in this way. However, with the coming of the deep learning era, the time consuming feature engineering is largely saved. Advanced architecture with meaningful language representation nowadays has better ability to capture the latent semantics in large-scale text data. Hence, there is a trend that more researchers are attempting to use deep neural networks to train a better text summarizer.

\subsection{Supervised Query-Based Summarization}
\begin{figure*}
    \centering
    \includegraphics[scale=0.5]{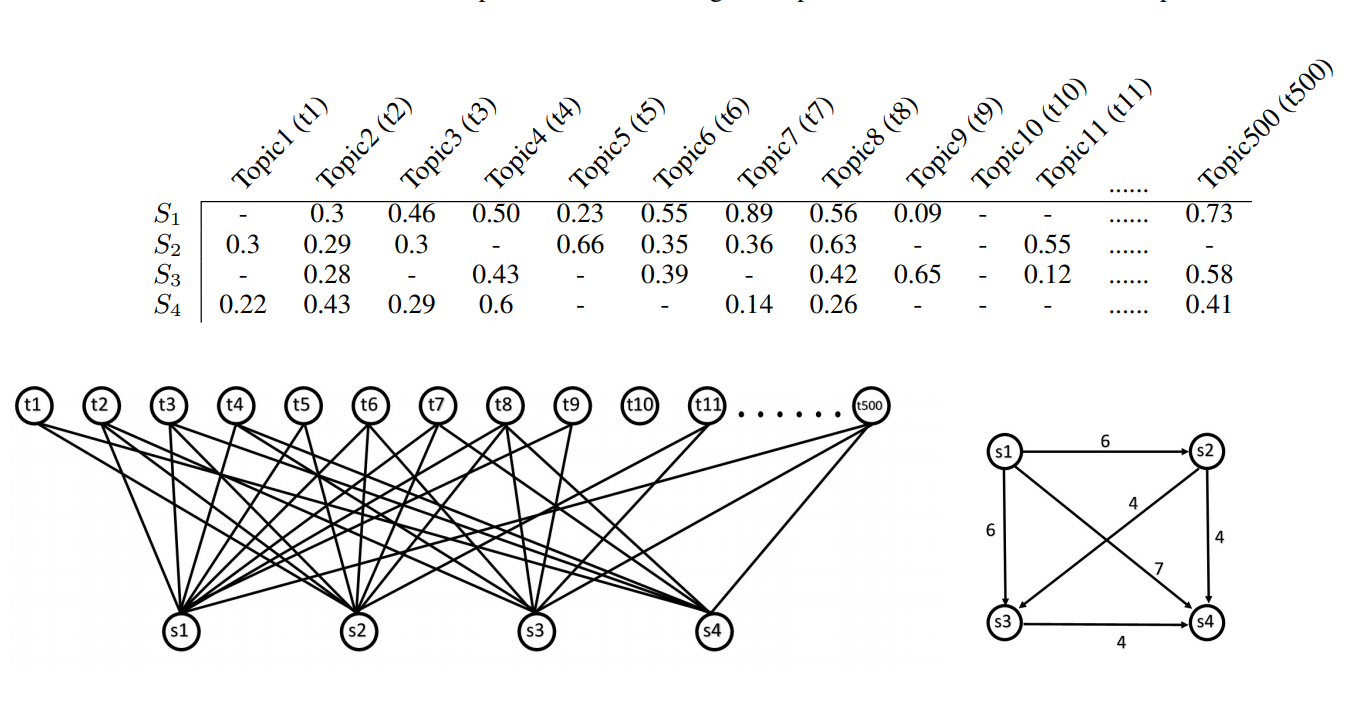}
    \caption{Sample Topical Graph in Parveen et al.\cite{parveen2015topical}}
    \label{fig:my_label}
\end{figure*}
In general, there are more supervised extractive text summarization than abstractive text summarization, yet few of the studies are found delving into query-based extractive summarization. Also, it is an interesting phenomenon that most query-based extractive summarization is performed in supervised learning One good guess might be that the lack of the dataset. Currently the most popular dataset for query-based text summarization is the DUC dataset. However, the amount of data is negligible compared to those large-scale datasets such as the NYT dataset, CNN/DM dataset, etc. But those datasets usually have only one general abstract of the passage, which is not suitable for query-based summarization tasks. Furthermore, the most prevalent metric used at present is the ROUGE score. Whether the ROUGE score is an effective metric in most text summarization tasks remains debated. In some way, the extracted text units from the source text might not be able to match the abstract if there are many words in the abstract that are not from the source text. With that being said, in this section, we will still discuss existing supervised generic extractive text summarization papers.

Parveen et al. (2015)\cite{parveen2015topical} implemented a graph-based text summarization method. The basic idea is to use the graph to represent the documents. As Parveen et al. suggested, their bipartite graph G = $(V_s, V_t, E_{t, s})$ has two sets of nodes connected by one set of edges. $V_s$ represents sentences and $V_t$ represents topics. The edge weight is the logarithmic sum of probabilities of words in topic t when there are more than one words of sentences present in topic t. The sample graph is shown in \textbf{Figure 1}.

As the common practice in other extractive text summarization methods, sentence ranking is again applied in their method. They used the HITS (Hyperlink Induced Topic Search)\cite{kleinberg1999authoritative} as the backbone algorithm for ranking sentences by importance. The initialization of the topic rank $Rank_{t_i}$ is 1 and the sentence rank $Rank_{s_i} = 1 + sim(s_i, title)$, where the similarity function is the cosine similarity and the title in the sentence rank is the title of the article.

To ensure the coherence of the summary, they proposed the coherence measure as:
\begin{align*}
    weighted_coh(s_i, P) &= weighted_Outdegree(s_i, P)\\
    norm_weighted_coh(s_i, P) &= 
    \frac{weighted_coh(s_i, P))}{\summation{}{}weighted_coh(s_i, P)}
\end{align*}
where $weighted_coh(s_i, P)$ is the outdegree of every sentence from the weighted projection graph and it is then normalized. The objective function to be optimized is defined as follow:
\begin{align*}
    Objective function = \underset{X, Y}{max}(f_i(X) + f_c(X) + f_t(Y))
\end{align*}
where X is a variable for sentences which contains boolean variables $x_i$ and Y is a variable for topics which contains boolean variables $y_j$. $f_i(X), f_c(X), f_t(Y)$ are separately corresponding to the sentence importance, coherence, and topic coverage. By solving the integer linear programming (ILP) optimization problem, the summaries are derived.

There are also studies about using BERT\cite{devlin2018bert} to perform extractive generic text summarization. In Liu et al. (2019)\cite{liu2019fine}, BERT is the backbone of the whole model and unsurprisingly, the model fine-tuned the BERT pre-trained model and achieved state-of-the-art on CNN/DM dataset.

As the \textbf{Figure 2} shown, three functional embeddings are stacked together before feeding into the BERT pre-trained model. After the representations are encoded by the BERT, the final summarization layers will fine-tune the BERT to make good extractions. The output sentence score will be forced to interval (0, 1) by sigmoid function for the simple classifier, which can be interpreted as the probability of the sentence to be in the extractive summarization. Apart from the simple fine-tune output layer, they also proposed an inter-sentence transformer to help the model focus on extracting document-level features:
\begin{align*}
    \Tilde{h}^l &= LN(h^{l-1} + MHAtt()h^{l-1})\\
    h^l &= LN(\Tilde{h}^l + FFN(\Tilde{h}^l))
\end{align*}
where $h^0$ would be the positional embedding of the sentence vector output by the BERT. MHAtt is the multi-head attention proposed by Vaswani et al.\cite{vaswani2017attention} and the superscript is the index of stacked transformer layer.

According to the synthesized papers, it is not hard to realize that the current extractive text summarization method is still relying on the sentence score. Also, graph-based is still prevalent in extractive text summarization but the general trend is more and more prone to apply large deep neural networks. It is promising that graph-based and deep neural networks can be incorporated together. There might be existing work on this but this survey failed to cover that.
\begin{figure*}
    \centering
    \includegraphics[scale=0.8]{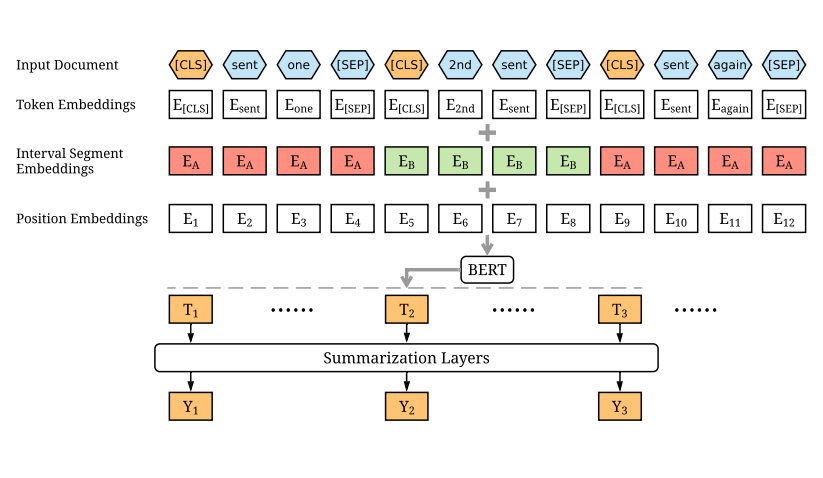}
    \caption{Fine-tune BERT Architecture in Liu et al.\cite{liu2019fine}}
    \label{fig:my_label}
\end{figure*}

\section{ABSTRACTIVE TECHNIQUES}
\subsection{Unsupervised Query-Based Summarization}
Interestingly, to the best of my knowledge, there is not yet any work in unsupervised query-based abstractive text summarization, which means previous studies about unsupervised query-based text summarization are all focused on the extractive summarization. However, there are many existing generic text summarization models using unsupervised abstractive methods.

Inspired by Rush et al. (2015)\cite{rush2015neural}, self-attention mechanism is becoming more and more prevalent in text summarization tasks, especially in abstractive text summarization tasks.

Li et al. (2017)\cite{li2017cascaded} observed that the important content is more likely to attract people's attention when they recall and digest what they have read for summaries. Based on this observation, they proposed a cascaded attention based unsupervised model for summarization purpose. The model is composed of two parts as suggested in the original paper. The first part is attention modeling for distillation, which distills the salient content of the documents. The attention modeling is subsequently composed of the reader, the recaller, and the cascaded attention modeling; The second part of the model is the compressive summary generation phase where fine-grained and coarse-grained sentence compression strategies are incorporated to produce compressive summaries.

In the distillation part, the reader is a LSTM or GRU network that imitates the reading process for a human reader who reads and forgets things at the same time. In this stage, the input text will be encoded into hidden states that will be further utilized in the recaller part. In the recaller part, the operation is similar but the output dimension is largely compressed to represent the condensed text which is the essence of the text summarization. The condensed vectors in the recaller are much smaller than that in the encoded hidden states. This process imitates the process of human readers recalling the content they read. People usually can only repeat the main idea of what they have read rather than all the details. For the cascaded attention part, similar to Bahdanau et al. (2015)\cite{bahdanau2014neural}, they first compute an attention vector, which represents the alignment weight of one sentence to another, using softmax as follow:
\begin{align*}
    a_{t, i}^h = \frac{exp(score(h_t^o, h_i^v))}{\summation{i'}{}exp(score(h_t^o, h_{i'}^v))}
\end{align*}
where $h_t^o$ is the output hidden state, and $h_i^v$ is the input hidden vector. The score function $score(\cdot)$ here is a content-based function to capture the relation between the two vectors. Finally, the context vector $c_t^h$ is a linear combination of the attention vector and corresponding input hidden state:
\begin{align*}
    c_t^h = \summation{i'}{}a_{i', h}^th^v_{i'}
\end{align*}

Li et al. also suggested that the attention mechanism itself can reflect the salience of the text unit. Thus, in their unsupervised learning method, the attention matrix is applied as one significant scoring criterion. Apart from considering the word salience obtained from the neural attention model, linguistically-motivated rules are also applied in their coarse-grained sentence compression part. The final summary is derived by incorporating the previously obtained word salience from the neural attention model, the compressed sentences from the coarse-grained sentence compression, and phrase-based optimization for summary construction.

Chu et al.\cite{chu2019meansum} proposed an unsupervised abstractive summarization model. The objective of their work is to abstractedly summarize multiple documents in the same topic. The backbone of their model is composed of two pairs of the encoder and decoder. One pair of the encoder and decoder is used for the reconstruction of the reviews while the other pair is used as a summarization module. Also, for simplicity, in this survey we would also call the encoder and the decoder for the reconstruction purpose as the reconstruction module.

According to their architecture which is shown in \textbf{Figure 3}, the multiple reviews are input into the first encoder so that all the reviews $x_j$ are encoded into a real vector $z_j = \phi_E(x_j)$. The encoded reviews are then fed into both the decoder in the reconstruction module and the summarization module. The decoder in the reconstruction module then decodes the encoded reviews as the reconstructed reviews. After the reconstructed reviews are generated, the auto-encoder reconstruction loss can be computed through cross entropy loss with respect to the original reviews as follow:
\begin{align*}
    l(\{x_1, x_2, \dots, x_k\}, \phi_E, \phi_D) = \summation{j=1}{k}l_{cross\_entropy}(x_j, \phi_D(\phi_E(x_j)))
\end{align*}
Basically, the reconstruction part of the model helps to train an auto-encoder to encode the reviews into a meaningful hidden representation of the original reviews. Then the encoded reviews also go to the summarization module. Before generating the abstractive summarization, the arithmetic mean of the encoded reviews are computed as the hidden representation of the summary and then they sample the summary by the decoder in the summarization module. The encoder in the summarization module encode the generated summary back to the topic representation and the previous encoded reviews are again used to compute the similarity to the topic representation using the following formula proposed in this paper:
\begin{align*}
    l_{sim}(\{x_1, x_2, \dots, x_3\}, \phi_E, \phi_D) = \frac{1}{k}\summation{j=1}{k}d_{cos}(z_j, \Bar{z})
\end{align*}
where $\Bar{z}$ is the topic representation encoded by the auto-encoder in the summarization module, $z_j$ is the j-th encoded review, and the similarity distance is computed by cosine similarity.

\begin{figure*}
    \centering
    \includegraphics[scale=0.5]{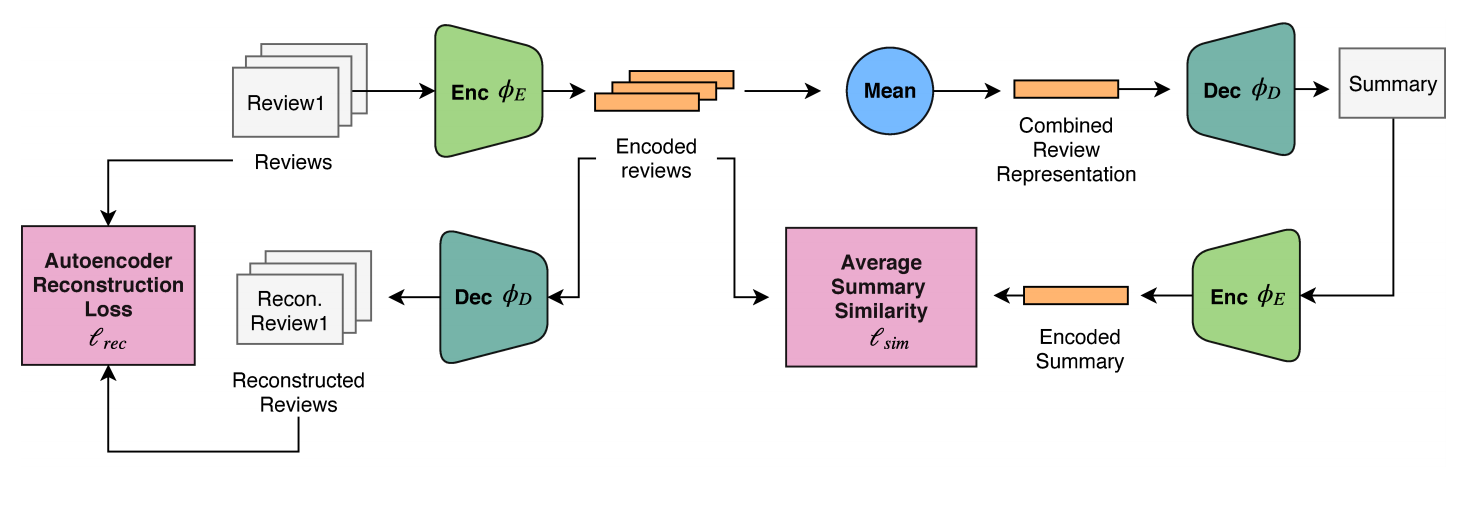}
    \caption{Architecture of MeanSuM model in Chu et al.\cite{chu2019meansum}}
    \label{fig:my_label}
\end{figure*}

By synthesizing two unsupervised abstractive generic text summarization, we find that both methods use a reconstruction circle to train the encoder and decoder to learn the semantic of the text. Also, the neural attention mechanism played a significant role of capturing the salient text content. Although related query-based text summarization work has not been found yet, the two paradigms indicate that the neural attention models have high potential in applying to query-based abstractive text summarization.

\subsection{Supervised Query-Based Summarization}
As discussed in previous sections, unsupervised and weakly supervised query-based abstractive text summarization methods are rarely studied. However, there are already many related works in supervised methods. Since abstractive text summarization methods typically require text generation, it is natural that most of the abstractive text summarization methods use the neural model, especially the neural attention model.

Before delving more into the query-based summarization, one paradigm of abstractive text summarization is worthwhile to mention first. In this survey, we have mentioned the pointer-generator abstractive summarization model\cite{see2017get} several times which was proposed by See et al. In their work, CNN/DM dataset is used to perform supervised text summarization. Their work made remarkable improvement on the liable factual details, self repetition, and out-of-vocabulary (OOV) issues that previous neural models have.

The overall model majorly contains two parts: the hybrid pointer-generator network and the coverage mechanism. The hybrid pointer-generator network is literally composed of the pointer and the generator. The pointer will point to some chosen word and copy the word from the source text while the generator generates new text to ensure the readabilities of the summaries. Before See et al., there are already existing prototypes of the generator in this summarization model. The baseline model from Nallapati et al. (2016)\cite{nallapati2016abstractive} is one of the prototype generators. The backbone of the baseline model is a probability distribution calculated by the self-attention mechanism proposed by Bahdanau et al.\cite{bahdanau2014neural} as we mentioned above. The objective of the baseline model is to minimize the negative log likelihood of the target word for each time step. The pointer network is derived from Vinyals et al. (2015)\cite{vinyals2015pointer}. By fusing the two networks together, the model compute a generation probability $p_{gen}$ as follow:
\begin{align*}
    p_{gen} = \sigma(w_{h^*}^Th^* + w_{s}^Ts_t + w_x^Tx_t + b_{ptr})
\end{align*}
where vectors $w_{h^*}, w_s, w_x$ and scalar $b_{ptr}$ are learnable parameters and $\sigma$ is the sigmoid function. $h^*_t, s_t, x_t$ are the context vector, the decoder vector, the decoder input vector at time step t separately. Since the generator has already computed the probability distribution among the vocabulary, and the attention vector has already been computed previously, those values next are linearly interpolated by the generation probability $p_{gen}$ using the formula:
\begin{align*}
    P(w) = p_{gen}P_{vocab}(w) + (1 - p_{gen})\summation{i:w_i = w}{}a_i^t
\end{align*}

To solve the repetition problem which is a common issue in sequence-to-sequence models, See et al. also incorporate the pointer-generator model with the coverage model proposed by Tu et al. (2016)\cite{tu2016modeling}. The coverage vector $c^t$ is the sum of attention distributions over all previous decoder time steps:
\begin{align*}
    c^t = \summation{t'=i}{t-1}a^{t'}_i
\end{align*}
With the introduction of the coverage vector, the attention mechanism introduced by Bahdanau et al. is modified to include the coverage vector term as follow:
\begin{align*}
    e_i^t &= v^Ttanh(W_hh_i + W_ss_t + w_cc_i^t + b_{attn})\\
    a^t &= softmax(e_i^t)
\end{align*}
By controlling the coverage loss, which is defined as follow:
\begin{align*}
    covloss = \summation{i}{}min(a_i^t, c_i^t)
\end{align*}
the penalty of repeatedly attending to the same locations will force to reduce the repetition of the covered content. The overall loss function for the model at t time step is the linear combination of the negative log likelihood of the pointer-generator model and the weighted coverage loss at t time step:
\begin{align*}
    loss_t = -logP(w_t^*) + \lambda\summation{i}{}min(a_i^t, c_i^t)
\end{align*}

As mentioned above, See et al. only focus on the generic abstractive text summarization, yet many works have been done in query-based abstractive text summarization. Hasselqvist et al. (2017)\cite{hasselqvist2017query} presented a model that generates summaries with respect to the input queries. However, the input query are the sentences rather than query words. Similar to the summarization model in See et al., this paper also applies the pointer-generator network for their summarizers. The structure of the pointer-generator network is shown in \textbf{Figure 4} which is a graph from the original paper. To manipulate the summary using the input query, a query auto-encoder is used to encode the input query. Also, there is a document encoder that encodes sequences in the document through time steps. In each time step for the document encoder, the output will go to the self-attention mechanism part for the pointer part. The encoded sequence in the last step also goes to the summary decoder accompanied by the encoded input query. By leveraging the information in both encoding vectors, the summary decoder will output the probability distribution over the whole vocabulary. This part, correspondingly, is called the generator in the pointer-generator network. Among these encoders and decoders, only the document encoder used bidirectional RNN, specifically GRU, while the other two applied unidirectional GRU.

The generator part of the model will produce a generation probability which is a softmax function over the output from two linear transformations on the decoder state and context vector $z_{tj}$. Specifically, the formula is:
\begin{align*}
    p_{tj}^{gen} = \frac{exp(z_{tj})}{\sum_kz_{tk}}
\end{align*}
The pointer probability at time step t is also computed using the formula:
\begin{align*}
    p_t^{ptr} = \sigma(v_{ptr}^T[s_t, E(y_{t-1}), c_t] + b_{ptr})
\end{align*}
The sigmoid function forces the output value to fall into the interval (0, 1), which is interpreted as the probability of copying a word from the source text. The final output is decided by the value of the pointer probability, which is sort of different from what See et al. did. When the pointer probability is greater than 0.5, then the final output copies the pointed word from the source text, otherwise the output would be generated according to the generating probability distribution.

This supervised model is finally trained by minimizing the defined training loss. The losses are from the pointer mechanism, the generator mechanism, and the attention mechanism. The overall loss is a simple summation of the losses from those three parts. The formulas are as follows:
\begin{align*}
    L_{ptr} &= \summation{t=1}{N_s}(x^{ptr}_t(-logp_t^{ptr}) + (1 - x_t^{ptr})(-log(1 - p_t^{ptr})))\\
    L_{gen} &= \summation{t=1}{N_s}(1 - x_t^{ptr})(-logP_t^{gen}(w^*))\\
    L_{att} &= \summation{t=1}{N_s}x_t^{ptr}(-log\alpha_{ti^*})\\
    L &= \frac{1}{N_s}(L_{gen} + L_{att} + L_{ptr})
\end{align*}
where $N_s$ is the length of the target summary, $w^*$ is the t-th target in the summary in the generation vocabulary, $i^*$ is the index in the input document to point to.

\begin{figure*}
    \centering
    \includegraphics[scale=0.5]{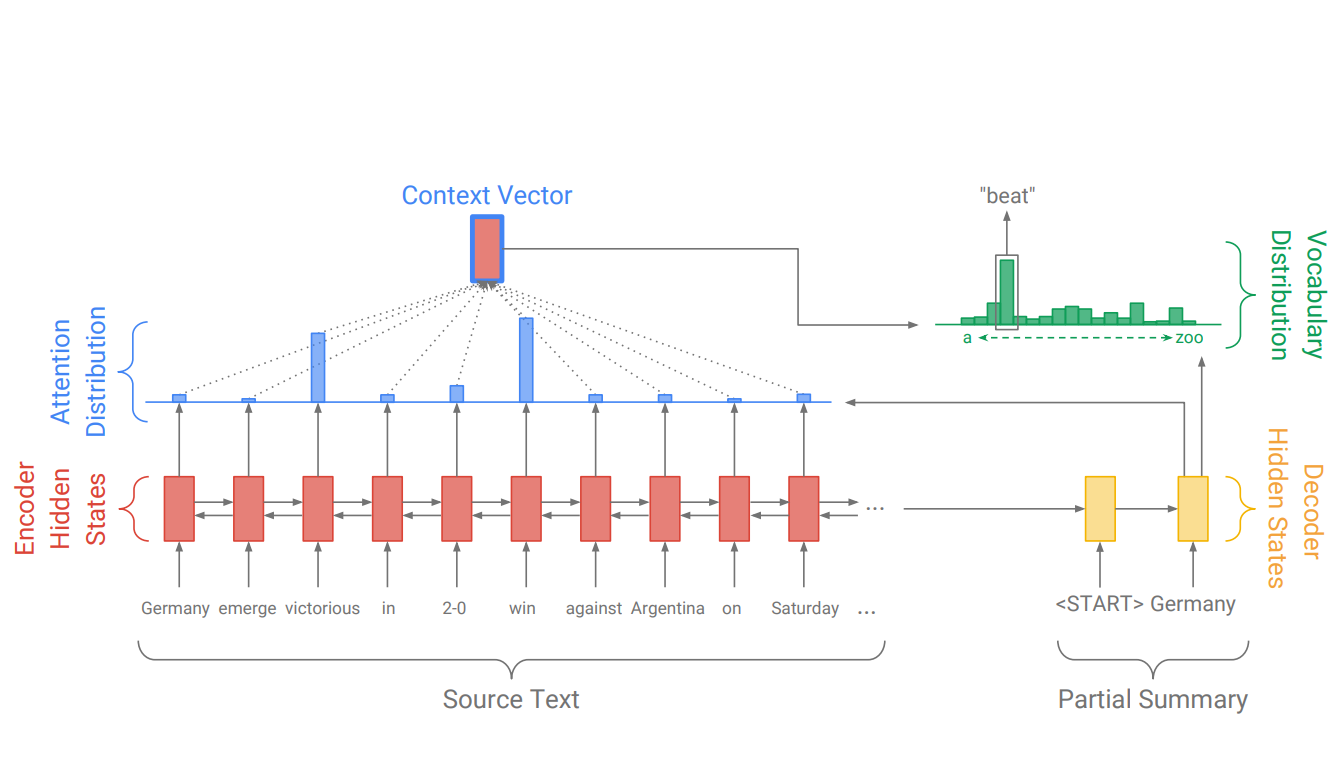}
    \caption{Architecture of Pointer-generator Network in See et al.\cite{see2017get}}
    \label{fig:my_label}
\end{figure*}

Although Hasselqvist et al.\cite{hasselqvist2017query} also applied the pointer-generator framework, from the details above, we can tell that the mechanisms they use are very different. However, another query-based abstractive text summarization work is much more similar to See et al. except for the repetition reduction mechanism.

Nema et al. (2017)\cite{nema2017diversity} introduced a diversity based attention model to tackle the query-based abstractive summarization task. The objective is to find a $\textbf{y}^*$ that maximize the probability $P(\textbf{y}|\textbf{q, d})$ where $\textbf{y}$ is the summary, $\textbf{q, d}$ are query and the document. The decomposed probability can be written as:
\begin{align*}
    y^* = \underset{y}{argmax}\pai{t=1}{m}p(y_t|y_1, y_2, \dots, y_{t-1}, \textbf{q, d})
\end{align*}
The proposed method has encoders for both the query and the document. Unlike that in Hasselqvist et al., the encoder for the query here is still unidirectional RNN while Hasselqvist et al. used bidirectional RNN. Another difference between their model lies in that Nema et al. applys attention mechanism for both the query and the document. The attention mechanism for the query is performed by:
\begin{align*}
    a_{t, i}^q &= v_q^Ttanh(W_qs_t + U_qh_i^q)\\
    \alpha_{t, i}^q &= \frac{exp(a_{t, i}^q)}{\sum_{j=1}^{k}exp(a_{t, j}^q)}\\
    q_t &= \summation{i=1}{k}\alpha_{t, i}^qh_i^q
\end{align*}
The $q_t$ is called the query representation. The attention mechanism for the document is similar to the attention mechanism for the query but it will also incorporate with the query representation stated above:
\begin{align*}
    a_{t, i}^d &= v_d^Ttanh(W_ds_t + U_dh_i^q + Zq_t)\\
    \alpha_{t, i}^d &= \frac{exp(a_{t, i}^d)}{\sum_{j=1}^{k}exp(a_{t, j}^d)}\\
    d_t &= \summation{i=1}{k}\alpha_{t, i}^dh_i^d
\end{align*}

See et al. proposed the coverage mechanism to leverage the covered information and reduce the self repetition issue. In Nema et al., they proposed the diversity based attention model with the hypothesis that the context vector being fed into the decoder at the consecutive time might be similar if the decoder produces the same phrase/word multiple times. With that being said, the paper experiments on four models ($D_1, D_2, SD_1, SD_2$) to solve the problem.

In concise, $D_1$ is to put a hard orthogonality constraint on the context vector $d'_t$ as:
\begin{align*}
    d'_t = d_t - \frac{d_t^Td'_{t-1}}{d^T_{t-1}d'_{t-1}}d'_{t-1}
\end{align*}
and $SD_1$ is the relaxed version of $D_1$ with a gating parameter $\gamma_t$:
\begin{align*}
    \gamma_t &= W_gd_{t-1} + b_g\\
    d'_t &= d_t - \gamma\frac{d_t^Td'_{t-1}}{d^T_{t-1}d'_{t-1}}d'_{t-1}
\end{align*}
Both $D_1$ and $SD_1$ models ensure the context vector to be diverse with respect to the previous context vector one time step ahead whereas the mechanism still lacks the consideration of the previous context vectors before time step t - 1.

In consideration of the previous information before time step t - 1, $D_2$ model modifies the LSTM cell to compute the new state at each time step:
\begin{align*}
    i_t &= \sigma(W_id_t + U_ih_{t-1} + b_i)\\
    f_t &= \sigma(W_fd_t + U_fh_{t-1} + b_f)\\
    o_t &= \sigma(W_od_t + U_oh_{t-1} + b_o)\\
    \hat{c}_t &= tanh(W_cd_t + U_ch_{t-1} + b_c)\\
    c_t &= i_t\odot\hat{c}_t + f_t\odot c_{t-1}\\
    c_t^{diverse} &= c_t - \frac{c_t^Tc_{t-1}}{c_{t-1}^Tc_{t-1}}c_{t-1}\\
    h_t &= o_t\odot tanh(c_t^{diverse})\\
    d'_t &= h_t
\end{align*}
The information gate $i_t$, forget gate $f_t$, and output gate $o_t$ are all derived from LSTM cells. Then for the diversity from the previous time step, the orthogonality transformation which is used in the $D_1$ model is performed again. Correspondingly, in $SD_2$ model, the modification is to change the hard orthogonality to a relaxed version with a parameter:
\begin{align*}
    g_t &= \sigma(W_gd_t + U_gh_{t-1} + b_o)\\
    c_t^diverse &= c_t - g_t\frac{c_t^Tc_{t-1}}{c_{t-1}^Tc_{t-1}}c_{t-1}
\end{align*}

Based on the synthesis in this section, we can find that recent study about abstractive text summarization typically applies the encoder-attention-decoder template. All three papers, no matter if it is generic text summarization or query-based text summarization, are the instantiation of this template. Although we have seen great achievements in this topic, the self repetition problem is still dazzling in this topic. It is still humble to predict that this template has yet not completely exploited and more methods with better self repetition reduction are of high potential.

\section{CONCLUSION}
This survey is intended to discuss query-based text summarization. The survey classifies the papers in both query-based summarization and generic summarization into four taxonomies in terms of the machine learning method (supervised or unsupervised), and the summary type (extractive or abstrative). The intention is to introduce some interesting query-based text summarization work but in some taxonomy the related work is rare or missing. The corresponding generic text summarization is also surveyed in each section to indicate the potential of transferring the method to the query-based summarization.

\bibliographystyle{ACM-Reference-Format}
\bibliography{survey}
\end{document}